\documentclass[floatfix,notitlepage,twocolumn]{revtex4-1}%
\usepackage{graphicx}%

\usepackage{amsmath}%
\usepackage{amsfonts}%
\usepackage{amssymb}
\usepackage{bm}
\usepackage{comment}
\usepackage{soul} 
\usepackage{color}
\usepackage{multirow}
\usepackage[breaklinks]{hyperref}

\def\ve{{\varepsilon}}
\def\k{{ {\bm k} }}

\def\0{{ {\bm 0} }}

\def\expo{{ {\rm e} }}
\def\hy{{ {\mathchar`-} }}
\def\ang{{ \mathrm{\mathring{A}} }}
\allowdisplaybreaks[4]

\allowdisplaybreaks[4]

\begin{document}
\title{Magnetic, transport and topological properties of Co-based shandite thin films}

\author{ 
Kazuki Nakazawa$^{1,2}$, Yasuyuki Kato$^1$ and Yukitoshi Motome$^1$} 

\date{\today }

\begin{abstract}
The kagome ferromagnet, Co-based shandite Co$_3$Sn$_2$S$_2$, shows a large anomalous Hall effect (AHE) associated with the Weyl nodes. A thin film with a Co kagome monolayer was predicted to exhibit the quantum AHE, which awaits the experimental realisation. However, it is challenging to precisely predict how the Weyl nodes reside in thin films where the lattice and electronic structures are in general different from the bulk. Here we report comprehensive {\it ab initio} results for thin films of Co$_3$Sn$_2$S$_2$ with one, two and three Co layers with Sn or S surface terminations. We find that all the Sn-end films stabilise a ferromagnetic state similar to the bulk, and retain the large AHE down to the monolayer limit where the AHE is quantised, while the magnetic and topological properties drastically change with the number of Co layers in the S-end films. Our results would stimulate further experimental exploration of thin Weyl materials. 
\end{abstract}

\address{
$^1$Department of Applied Physics, The University of Tokyo, Tokyo 113-8656, Japan
\\
$^2$RIKEN Center for Emergent Matter Science, 2-1 Hirosawa, Wako, Saitama 351-0198, Japan
}
 
\maketitle

\section*{Introduction} 

After the prediction of the Berezinskii-Kosterlitz-Thouless transition and Haldane's seminal work for a spontaneous quantum anomalous Hall effect (AHE) in two spatial dimensions, it has been widely recognised that condensed matters can be characterised using the concept of topology~\cite{Nobel}. The Weyl semimetals are a representative in three dimensions that have attracted much attention because of the variety of interesting physical properties arising from their unique topology~\cite{AMV}. The energy spectrum of the Weyl semimetals possesses Weyl nodes, which can be regarded as ``monopoles" with chirality $+1$ or $-1$ in momentum space. The Weyl nodes always appear in pairs with positive and negative chirality~\cite{NN}, and this feature generates the Fermi arc on the surface through the bulk-edge correspondence~\cite{Murakami}, as indeed observed by the angle-resolved photoemission spectroscopy~\cite{Xu, Souma}. The peculiar topology of the Weyl semimetals is also detected in quantum transport phenomena, such as the large AHE~\cite{Murakami,Burkov,Suzuki}, the anomalous Nernst effect (ANE)~\cite{Sakai} and the negative magnetoresistance associated with the chiral anomaly due to the monopoles~\cite{Burkov,XHuang,Zhang}. 

A prescription for the Weyl semimetals is to lift the degeneracy in the Dirac nodes by breaking symmetry. 
The Weyl semimetals with broken spatial inversion symmetry have been confirmed in several materials, e.g., transition metal pnictides~\cite{Xu, Souma,WFFBD} and dichalcogenides~\cite{Soluyanov,Feng}, whereas those with broken time-reversal symmetry have been rather scarce, with only a few theoretical proposals~\cite{WTVS,XWWDF,KF} and experiments under the magnetic field~\cite{Suzuki,Sakai}. Recently, however, a kagome ferromagnet Co-based shandite, Co$_3$Sn$_2$S$_2$, has been discovered to possess the Weyl nodes by spontaneous breaking of time-reversal symmetry, in the vicinity of $60$~meV from the Fermi level~\cite{QXu, Wang, Liu}. Indeed, theoretical and experimental studies revealed that the giant AHE and ANE observed in this compound are of the intrinsic contribution from the Weyl nodes~\cite{QXu,Wang,Liu,Guin,Yang,Minami,Yanagi}.

\begin{figure*}[t]
\begin{center}
 \includegraphics[width=140mm]{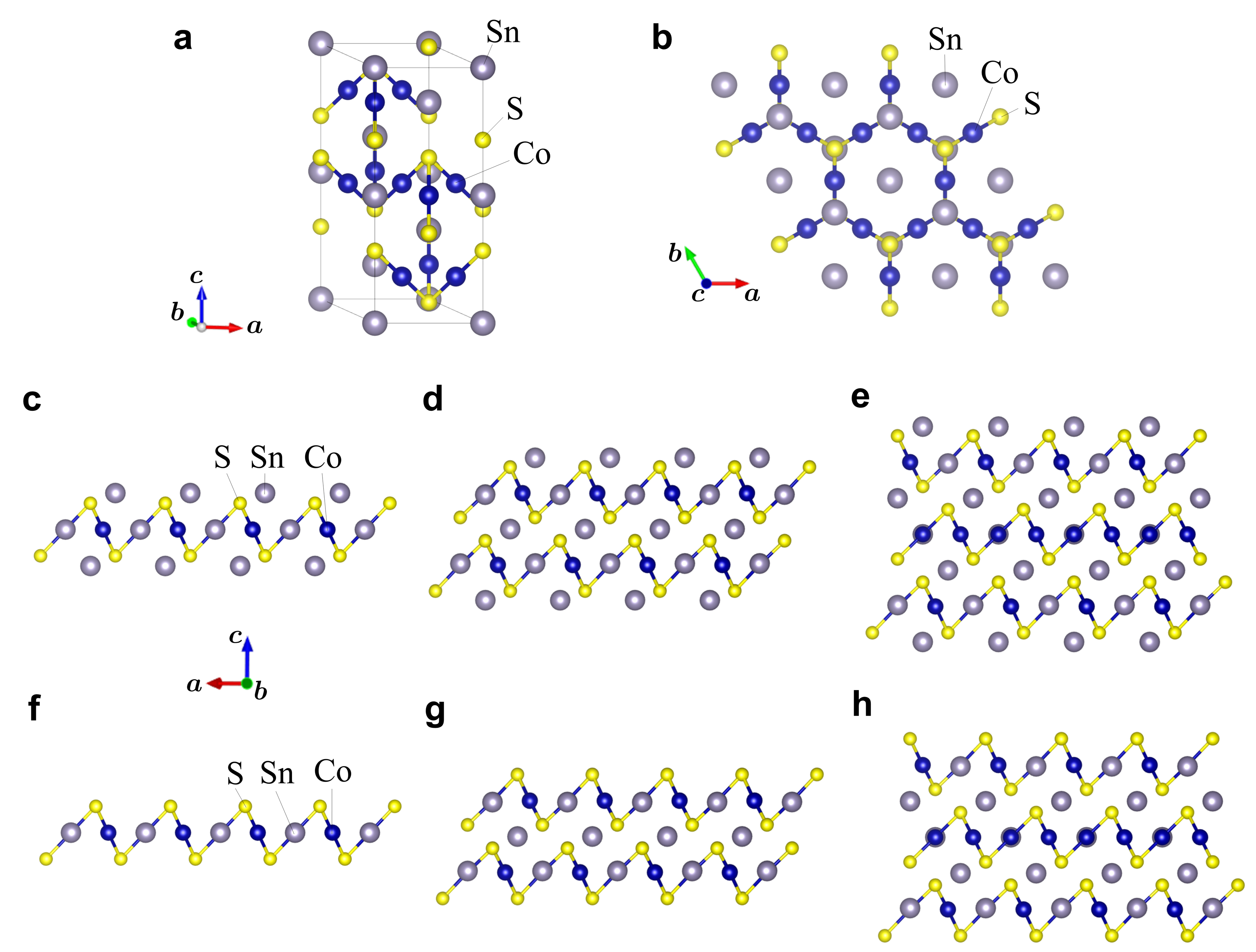}
\end{center}
\vspace{-5mm}
 \caption{
{\bf Lattice structures of the bulk and thin films of Co$_3$Sn$_2$S$_2$.} 
{\bf a,} Bulk structure. The black box represents the conventional unit cell which contains three unit cells. {\bf b,} Top view of a kagome plane composed of Co ions. {\bf c-e,} Side views of the monolayer, bilayer and trilayer systems with Sn terminations. {\bf f-h}, Corresponding figures with S terminations. The lattice structures are visualised by VESTA~\cite{MI}. 
 }
\label{fig:Lattice}
\end{figure*}

Co$_3$Sn$_2$S$_2$ has a quasi-two-dimensional lattice structure with Co kagome layers, as shown in Fig.~\ref{fig:Lattice}a, b. Recently, two types of Co kagome monolayers with different surface terminations, Co$_3$Sn$_3$S$_2$ and Co$_3$SnS$_2$, were investigated theoretically~\cite{Muechler}. As a remnant of the Weyl nodes in the bulk, both monolayers are predicted to show a large AHE; especially in Co$_3$Sn$_3$S$_2$ with Sn-end surfaces, the Weyl gap opens at the Fermi level, which leads to the quantum AHE. Stimulated by this theoretical prediction, experimental attempts to fabricate monolayer systems have been carried out, and the magnetism, transport phenomena and surface states have been discussed for atomically thin films of Co$_3$Sn$_2$S$_2$~\cite{Fujiwara,Shiogai,Ikeda1,Ikeda2}. In the theoretical study, however, the lattice parameters were fixed at the bulk values for both monolayers, despite the fact that the lattice structures are expected to be relaxed in a different manner depending on the surface termination, sensitively affecting the magnetism and band topology of the atomically thin films~\cite{GG,BHuang}. More importantly, since the Weyl points are the topological singularities predicted for the three-dimensional bulk systems, it remains elusive how they appear while changing the film thickness. 

In this study, we perform comprehensive {\it ab initio} calculations of the band topology and transport properties for thin films of Co$_3$Sn$_2$S$_2$, with special attention to the optimal lattice structures and stable magnetic states. We study monolayers, bilayers and trilayers of Co kagome planes with two different surface terminations by Sn and S atoms (Fig.~\ref{fig:Lattice}c--h). From the obtained magnetic and electronic states, we compute the anomalous Hall conductivity and the anomalous Nernst conductivity. Through the systematic study in comparison with the bulk, we discuss the evolution of the topological properties with the number of kagome layers from one, two, three to infinity. 

\section*{Results and discussions}

\noindent
{\bf Lattice structures and magnetic states.} First, we optimise the lattice structures of each system (see Methods). The optimised structures are summarised in Table~\ref{tab:opt} (see also Supplementary Note 1). All the thin films maintain nearly the $p\bar{3}m1$ symmetry. Compared to the bulk, the in-plane lattice constants change by less than $\sim 0.6$~\% for the Sn-end systems, while substantial decreases are observed for the S-end systems, maximally about $3$~\% for the monolayer case. Such lattice deformation was not taken into account in the previous study for the monolayers~\cite{Muechler}. We note that the distance between the Co layers for bilayers and trilayers does not change so much from the bulk value, while it is slightly larger for the Sn-end systems than the S-end ones. We also estimate the differences in the formation energies between Sn-end and S-end thin films in each layer number (see Methods). The estimates suggest that the Sn-end monolayer, bilayer and trilayer are more stable than the S-end counterparts by $-1.74$, $-1.60$ and $-1.55$~eV per formula unit, respectively.

\begin{table*}
\centering
\newlength{\height} 
\setlength{\height}{3mm}
\begin{center}
\fontsize{8pt}{11pt}\selectfont
	\begin{tabular}{cccccc} \hline 
\multicolumn{2}{c}{ System } 							&	Lattice constant $(\ang)$	&	Co layer distance $(\ang)$	& Magnetic state									& Co moment ($\mu_{\rm B}$)		  	\\ \hline \hline	
\multirow{4}{*}{Sn-end} 	& monolayer					&	5.325					&	-							& FM $\parallel c$								& 0.415								\\ 
						& bilayer					&	5.355					&	4.434						& FM $\parallel c$								& 0.372								\\ 	 
						& \multirow{2}{*}{trilayer}		&	\multirow{2}{*}{5.370}		&	\multirow{2}{*}{4.436}  		& \multirow{2}{*}{FM $\parallel c$}					& 0.365 (top, bottom)  				\\  
						& 							&							& 								& 												& 0.362 (middle)  					\\ \hline
\multirow{4}{*}{S-end} 	& monolayer 				&	5.194					&	-							& FM $\parallel c$								& 1.01								\\ 	 	
						& bilayer					&	5.276					&	4.410						& interlayer AFM $\parallel b'$					& 0.339								\\ 	  
						& \multirow{2}{*}{trilayer}		&	\multirow{2}{*}{5.310}		&	\multirow{2}{*}{4.412}  		& \multirow{2}{*}{interlayer ferri $\parallel c$}		& 0.207 (top, bottom)					\\  
						& 							& 							& 								& 												& 0.080 (middle) 						\\ \hline	  
\multicolumn{2}{c}{\multirow{2}{*}{bulk}}					&	5.358 $(a,b)$ 			&	\multirow{2}{*}{4.403} 	 		& \multirow{2}{*}{FM $\parallel c$}					& \multirow{2}{*}{0.350} 				\\ 
						& 							&    13.123 $(c)$  			& 								& 												& 									\\ \hline
	\end{tabular}
	\\
\vspace*{2mm}
\caption{
{\bf Optimized lattice structures and stable magnetic states.} 
The lattice structure for the bulk is taken from the experimental data~\cite{Li}, whose in-plane lattice constants and the Co layer distance are used for the initial guess for the optimization of the thin films. $a$, $b$ and $c$ denote the crystallographic axes defined in Fig.~\ref{fig:Lattice}. The structural optimizations are performed for a FM state, except for the S-end bilayer and trilayer where we assume interlayer AFM and ferrimagnetic states, respectively, by nonrelativistic {\it ab initio} calculations. For all the cases, the $p\bar{3}m1$ symmetry is almost preserved, while the lattice constant is considerably shortened in the S-end cases. All the Sn-end systems exhibit out-of-plane FM, whose Co magnetic moments approach the bulk value while increasing the layer number. In contrast, the S-end systems show different magnetic states depending on the layer number: out-of-plane FM for monolayer, interlayer AFM with in-plane FM moment directed to ${\bm b}' = {\bm b} - {\bm a}/2$ in each layer for bilayer, and interlayer ferrimagnetic state with out-of-plane FM moment in each layer for trilayer, where $\bm a$ and $\bm b$ are the in-plane lattice vectors. See also Fig.~\ref{fig:Sn_end}a, e, i, m and Fig.~\ref{fig:S_end}a, e, i.  
}
\label{tab:opt}
\end{center}
\end{table*}

\begin{figure*}
\vspace{-10mm}
\begin{center}
  \includegraphics[width=130mm]{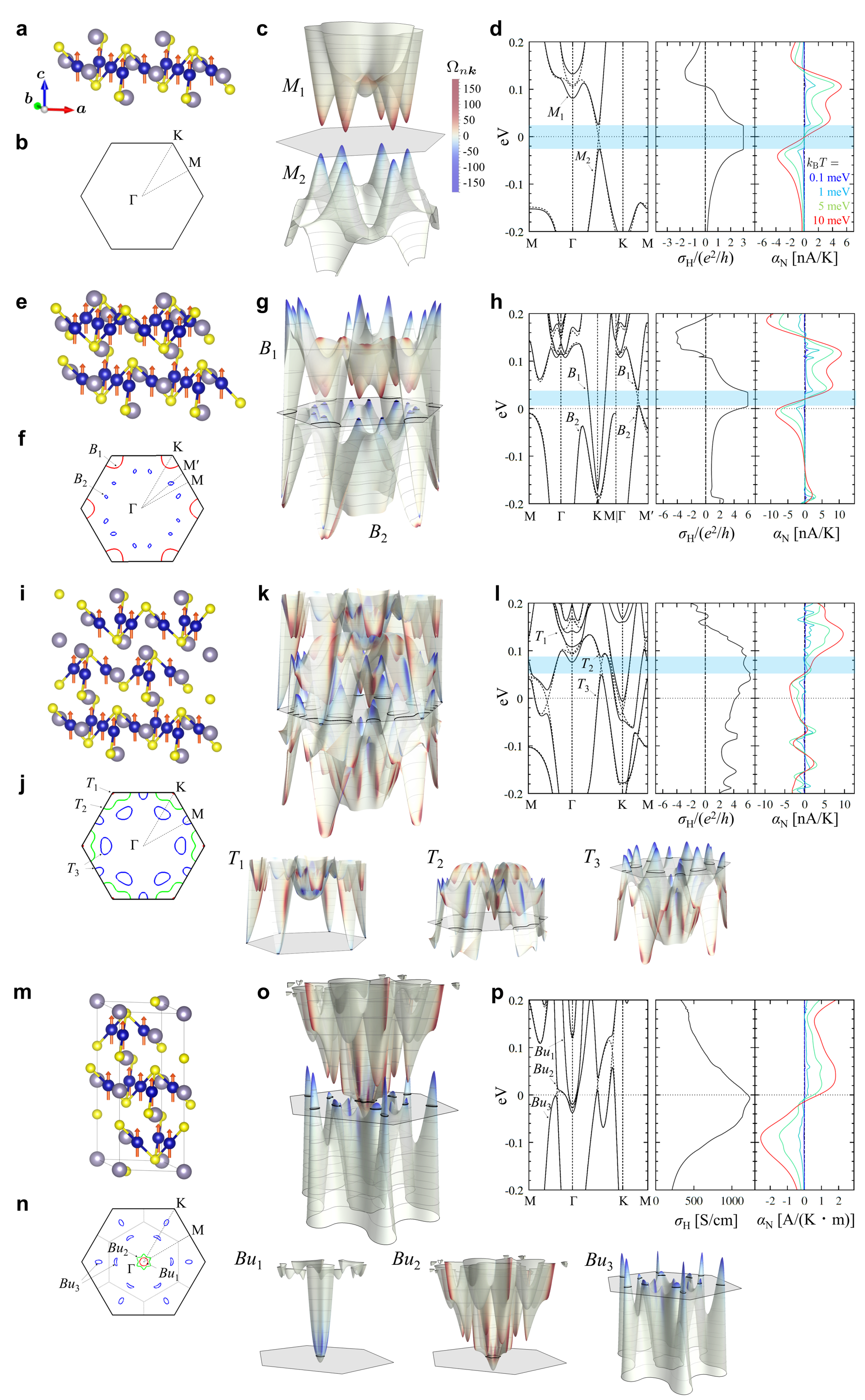}
\end{center}
\vspace{-5mm}
 \caption{
{\bf 
Magnetic states, Fermi surfaces, band structures with the Berry curvature $\Omega_{n \k}$, the anomalous Hall conductivity $\sigma_{\rm H}$ and the anomalous Nernst conductivity $\alpha_{\rm N}$ for the Sn-end films}: 
{\bf a-d,} monolayer, {\bf e-h,} bilayer and {\bf i-l,} trilayer. For comparison, we present the data for bulk in {\bf m-p}. The orange arrows in {\bf a, e, i} and {\bf m} represent the magnetic moments on the Co ions. The black or gray hexagons in {\bf b, f} and {\bf j} as well as {\bf c, g} and {\bf k} are the first Brillouin zone, while those in {\bf n} and {\bf o} depict the $k_z=0$ cut of the Brillouin zone of the bulk. M, M$'$, $\Gamma$ and K are the specific momenta in the first Brillouin zone and $M_n, B_n, T_n$ and $Bu_n$ denote the band labels near the Fermi level set at $0$~eV for monolayer, bilayer, trilayer and bulk, respectively. The band structures obtained from the fully relativistic and nonrelativistic calculations are represented by solid and broken lines, respectively, and $\sigma_{\rm H}$ at zero temperature $T=0$ and $\alpha_{\rm N}$ at the several temperatures are plotted in {\bf d, h, l} and {\bf p}, where $e, h$ and $k_{\rm B}$ are the elementary charge, Planck constant and Boltzmann constant, respectively. An out-of-plane ferromagnetic state is obtained throughout this Sn-end series, The monolayer is a Chern insulator with the Weyl gap between the bands $M_1$ and $M_2$, resulting to the quantised $\sigma_{\rm H}=3e^2/h$. The bilayer is metallic, while having a similar Weyl gap between the bands $B_1$ and $B_2$ above the Fermi level, leading to the plateau (unquantised) $\sigma_{\rm H} \simeq 6e^2/h$. The trilayer hosts electron and hole pockets and Weyl gaps, similar to the bulk. $\sigma_{\rm H}$ is always positive at Fermi level and shows similar energy dependence in this series. 
}
\label{fig:Sn_end}
\end{figure*}

\begin{figure*}
\vspace{-0mm}
\begin{center}
  \includegraphics[width=130mm]{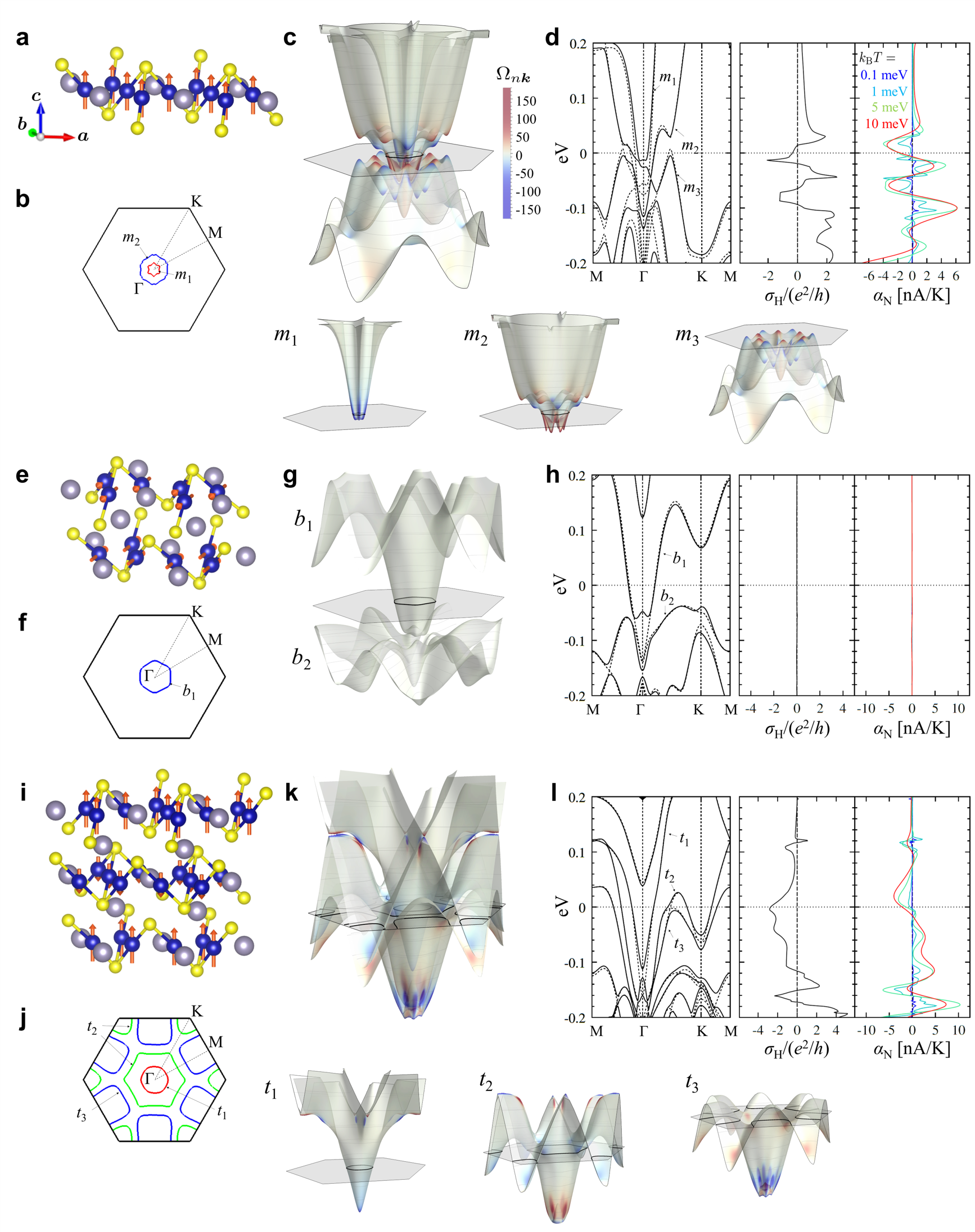}
\end{center}
\vspace{-5mm}
 \caption{
 {\bf  Magnetic states, Fermi surfaces, band structures with the Berry curvature $\Omega_{n\k}$, the anomalous Hall conductivity $\sigma_{\rm H}$ and the anomalous Nernst conductivity $\alpha_{\rm N}$ for the S-end films}: 
{\bf a-d,} monolayer, {\bf e-h,} bilayer and {\bf i-l,} trilayer. The notations are similar to those of Fig.~\ref{fig:Sn_end}. In this series, the stable magnetic states change with the number of layers: out-of-plane ferromagnetic, in-plane interlayer antiferromagnetic and out-of-plane interlayer ferrimagnetic states for the monolayer, bilayer and trilayer, respectively. While all the cases are metallic, the Berry curvature and the transport properties do not behave systematically with respect to the layer number. In particular, both $\sigma_{\rm H}$ and $\alpha_{\rm N}$ vanish in the bilayer case because of the band degeneracy. 
}
\label{fig:S_end}
\end{figure*}

\begin{figure*}
\begin{center}
\hspace*{-0mm}
 \includegraphics[width=120mm]{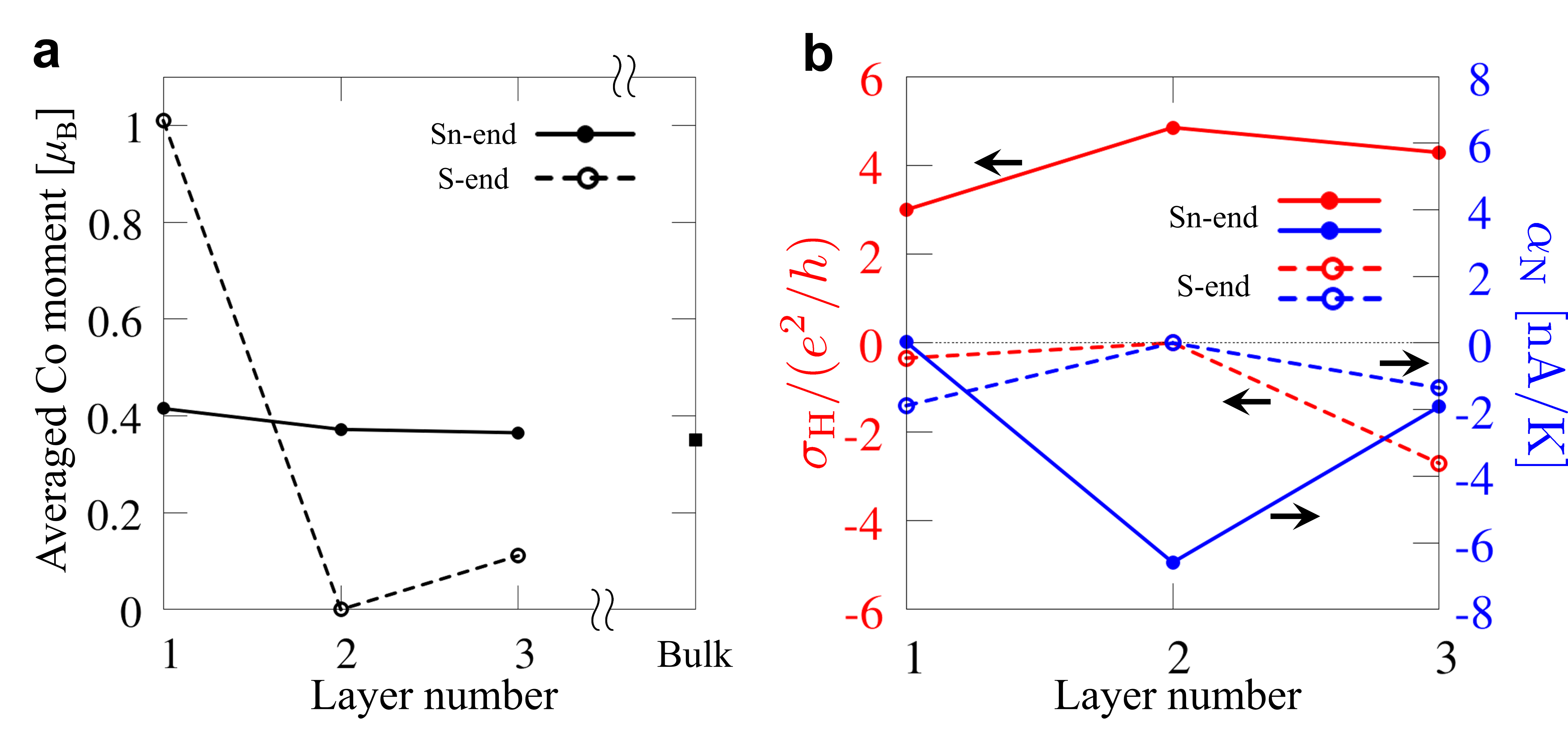}
\end{center}
\vspace*{-5mm}
 \caption{
{\bf Layer number dependence of the magnetic and transport properties of the thin films.} 
{\bf a,} The averaged magnetic moment per Co ion. In the Sn-end films where the out-of-plane ferromagnetic state is stable, the magnetic moment changes smoothly to the bulk value. In contrast, in the S-end cases, it shows nonmonotonic behaviour, reflecting the different magnetic states depending on the layer number. {\bf b,} The anomalous Hall conductivity $\sigma_{\rm H}$ at zero temperature and the anomalous Nernst conductivity $\alpha_{\rm N}$ at $k_{\rm B}T=5$~meV. $\sigma_{\rm H}$, which reflects the band topology through the Berry phase, has large positive values in the Sn-end series, whereas small negative values for monolayer and trilayer, and zero in bilayer in the S-end series. $\alpha_{\rm N}$ also shows contrasting behaviours between the two cases. 
}
\label{fig:summary}
\end{figure*}

Next, for the optimised structures, we compare the {\it ab initio} energy for various magnetic states (see Supplementary Note 2). The lowest-energy states and the magnitudes of Co moments are summarised in Table~\ref{tab:opt} (see Methods); see also Fig.~\ref{fig:Sn_end}a, e, i, m and Fig.~\ref{fig:S_end}a, e, i. For the bulk, a ferromagnetic state is obtained with Co moment of $0.350$~$\mu_{\rm B}$ along the out-of-plane direction, where $\mu_{\rm B}$ is the Bohr magneton; the total magnetic moment per unit cell is $1.00$~$\mu_{\rm B}$ because of small but nonzero antiparallel magnetic moments at Sn and S atoms. The result agrees well with the previous studies~\cite{Liu,Wang,Yanagi}. For the Sn-end thin films, similar out-of-plane ferromagnetic states are obtained for all the monolayer, bilayer and trilayer systems. The Co moment changes with the number of Co layers: $0.415$~$\mu_{\rm B}$ for monolayer, $0.372$~$\mu_{\rm B}$ for bilayer, and $0.365$~$\mu_{\rm B}$ (top and bottom layers) and $0.362$~$\mu_{\rm B}$ (middle layer) for trilayer; the value is monotonically reduced to the bulk one as the number of Co layers increases. We note that the net moment per primitive unit cell is $1.00$, $2.01$ and $3.00$~$\mu_{\rm B}$ for monolayer, bilayer and trilayer, respectively, indicating that the value per Co layer is the same as in the bulk. In contrast, for the S-end thin films, we obtain different magnetic states depending on the layer number: an out-of-plane ferromagnetic state for monolayer, an intralayer in-plane ferromagnetic and interlayer antiferromagnetic state for bilayer, and an intralayer out-of-plane ferromagnetic and interlayer ferrimagnetic state for trilayer, as shown in Fig.~\ref{fig:S_end}a, e, i, whose net magnetic moments per unit cell are $3.01$, $0.00$ and $1.00$~$\mu_{\rm B}$, respectively. The Co moment is also largely different from the Sn-end cases: $1.01$~$\mu_{\rm B}$ for monolayer, $0.339$~$\mu_{\rm B}$ for bilayer, and $0.207$~$\mu_{\rm B}$ (top and bottom layers) and $0.080$~$\mu_{\rm B}$ (middle layer) for trilayer. The contrasting behaviours between the Sn- and S-end series are summarised in Fig.~\ref{fig:summary}a. 

Our results indicate that the ferromagnetic order within each Co kagome layer is stable, regardless of the number of Co layers and the surface termination. In addition, the magnetic order has out-of-plane anisotropy, except for the S-end bilayer system. These suggest that the out-of-plane ferromagnetic state of the Co layer, which is a building block for the bulk system, is robust in most cases of thin films. Meanwhile, our results reveal that the surface termination matters to the interlayer magnetism as well as the magnetic anisotropy. Analysing the maximally-localised Wannier functions, we indeed find that the spatial extensions of the Co $d$ orbitals are substantially different for different surface terminations, which would be relevant to the interlayer magnetic interactions (see Supplementary Note 3). 

\noindent
{\bf Electronic structures and band topology.} The differences in the lattice structures and the stable magnetic states result in qualitatively different electronic states between the Sn- and S-end series. We show the Fermi surfaces in Fig.~\ref{fig:Sn_end}b, f, j, n and Fig.~\ref{fig:S_end}b, f, j and the band structures near the Fermi level in Fig.~\ref{fig:Sn_end}c, g, k, o and Fig.~\ref{fig:S_end}c, g, k. In the band structures, we also plot the Berry curvature for each band by colour which is computed for the tight-binding models constructed by the Wannier functions obtained from the {\it ab initio} calculations; we calculate the Chern number of each band as well (see Methods).

Let us first discuss the Sn-end films in Fig.~\ref{fig:Sn_end}. In the case of the monolayer, the band structure has a gap of $\sim 50$~meV at the Fermi level between the bands $M_1$ and $M_2$, and there is no Fermi surface. The bands $M_1$ and $M_2$ have the Chern number $6$ and $-3$, respectively. The gap is opened by the spin-orbit coupling at the Weyl nodes on the $\Gamma$-$\rm K$ line in momentum space, resulting in the large values of the Berry curvatures near the band top and bottom. Thus, the system is an insulator with the Weyl gap, as pointed out in the previous study without structural optimisation~\cite{Muechler}. This is reasonable, given the little changes in the lattice structure as well as the magnetism after the optimisation in our calculations. In the bilayer, however, the system becomes metallic with the electron Fermi surfaces around the $\rm K$ point in the band $B_1$ and the pairs of the hole pockets on both sides of each $\Gamma$-$\rm M$ line in the band $B_2$. The Chern numbers of the bands $B_1$ and $B_2$ are $9$ and $-1$, respectively. By closely looking at the band structure, we can identify the Weyl gap of $\sim 30$~meV between $B_1$ and $B_2$ slightly above the Fermi level, which appears to correspond to that in the monolayer case but split into two, leading to the hole pocket pair; see also the band structure in Fig.~\ref{fig:Sn_end}h. In the trilayer system, the band structure becomes more complicated, but still retains the characteristic features in the monolayer and bilayer: One of the Weyl gaps opens between the bands $T_2$ and $T_3$ on the $\Gamma$-K line above the Fermi level, while the other two are along the $\Gamma$-M line, as shown in the band structure in Fig.~\ref{fig:Sn_end}l. The Chern numbers of the bands $T_1$, $T_2$ and $T_3$ are $-3$, $6$ and $3$, respectively. These overall systematic changes of the electronic states appear to smoothly connect to the bulk. In fact, as shown in Fig.~\ref{fig:Sn_end}n, o, the $k_z=0$ cut of the bulk band structure shows similar distribution of the Weyl gap and the Berry curvature to the above Sn-end films. In particular, the large Berry curvature in the Sn-end films can be regarded as a remnant of the intersections of the nodal gap in the bulk system~\cite{QXu,Minami}. Thus, in the Sn-end series, where the out-of-plane ferromagnetic state persists, the electronic structure and the topological properties change systematically as the number of Co layers increases. 

Next, we discuss the S-end cases. In the monolayer, the system is metallic with two electron pockets around the $\Gamma$ point as shown in Fig.~\ref{fig:S_end}b, in contrast to the Sn-end case. The result is also different from the previous study without structural optimisation~\cite{Muechler}. The difference is ascribed to the substantial shrinkage of the lattice constants shown in Table~\ref{tab:opt}. We find considerable values of the Berry curvature in the bands close to the Fermi level, which affect the transport properties as discussed later. In the case of the bilayer, the system is also metallic with one electron pocket around the $\Gamma$ point; all the bands are doubly degenerate due to the interlayer antiferromagnetic ordering, cancelling out the Berry curvature. In the trilayer, there are three bands forming two electron pockets around the $\Gamma$ point, one electron pocket around the K point, and one hole pocket around the M point. In each band, the Berry curvature takes nonzero values, similar to the monolayer case. In addition, we find two small Weyl gaps between the bands $t_1$ and $t_2$ along the $\rm M$-$\Gamma$ line and another one between $t_2$ and $t_3$ along $\Gamma$-$\rm K$ line close to the Fermi level. Although this apparently resembles the case of Sn-end trilayer and bulk systems, these bands show qualitatively different Berry curvatures due to the different magnetic state. Thus, in stark contrast to the Sn-end cases, the S-end series shows no systematic change not only in the magnetic orders but also in the electronic structure and the band topology.

\noindent
{\bf Transport properties.} Now we turn to the transport properties. First of all, in the bulk case, our results on the anomalous Hall conductivity $\sigma_{\rm H}$ and the anomalous Nernst conductivity $\alpha_{\rm N}$ well reproduce the previous ones~\cite{Liu,Wang,Yanagi}, as shown in Fig.~\ref{fig:Sn_end}p. Then, in the Sn-end series, the monolayer shows a quantised AHE with $\sigma_{\rm H}=3e^2/h$, where $e$ and $h$ are the elementary charge and the Planck constant, respectively, corresponding to the Weyl gap in the blue-shaded energy window in Fig.~\ref{fig:Sn_end}d including the Fermi level. The result coincides with the previous one without structural optimisation~\cite{Muechler}. In the bilayer, we find that $\sigma_{\rm H}$ exhibits a plateau at $\simeq 6e^2/h$ slightly above the Fermi level in the blue-shaded region, which corresponds to the Weyl gap between the bands $B_1$ and $B_2$, as shown in Fig.~\ref{fig:Sn_end}h. The plateau with almost twice $\sigma_{\rm H}$ of the monolayer case is due to the systematic change of the band structure from monolayer to bilayer mentioned above. Note that the contribution from the band $B_1$ crossing the Fermi level is negligible. Such a plateau is absent in the trilayer case, but a large value of $\sigma_{\rm H}$ is still observed in the energy window corresponding to the gap between the bands $T_2$ and $T_3$, as shown in Fig.~\ref{fig:Sn_end}l. Thus, in the Sn-end series, we obtain systematic evolution of the AHE, from the quantum AHE in the monolayer to the large AHE in the bulk limit, reflecting the band structure with the Weyl gap. We note that $\sigma_{\rm H}$ is always positive in the Sn-end series. The results are summarised in Fig.~\ref{fig:summary}b.  

In contrast, $\sigma_{\rm H}$ does not show systematic changes in the S-end cases, as shown in Fig.~\ref{fig:S_end}d, h and l: $\sigma_{\rm H}/(e^2/h) \simeq -0.34$, $0.00$ and $-2.71$ for the monolayer, bilayer and trilayer, respectively. The absence of the AHE in the bilayer case is due to the cancellation of the Berry curvature by the band degeneracy discussed above, and the negative values of $\sigma_{\rm H}$ for the monolayer and trilayer originate from the different band topology from the Sn-end series. Note that in the monolayer case, $\sigma_{\rm H}$ is largely different from that of the previous study~\cite{Muechler} because of the structural optimisation. 

We also compute the anomalous Nernst conductivity $\alpha_{\rm N}$ for both series of the films. According to the generalised Mott formula $\alpha_{\rm N} = -\frac{k_{\rm B}}{e} \int d\ve s(\ve) \frac{\partial \sigma_{\rm H}(\ve)}{\partial \ve}$~\cite{XYFN,Yanagi}, $\alpha_{\rm N}$ is related with the energy derivative of $\sigma_{\rm H}$. This holds for our results, as plotted in Fig.~\ref{fig:Sn_end}d, h, l, p and Fig.~\ref{fig:S_end}d, h, l. In the Sn-end series, $\alpha_{\rm N}$ is almost zero at the Fermi level for the monolayer because of the quantisation of $\sigma_{\rm H}$, while it takes a negative value for the bilayer and trilayer cases corresponding to the positive derivative of $\sigma_{\rm H}$. Interestingly, $\alpha_{\rm N}$ takes a large negative value for the bilayer since the Fermi level is on the verge of the plateau-like feature of $\sigma_{\rm H}$ where $\sigma_{\rm H}$ increases rapidly. Since $\alpha_{\rm N}$ is positive in the bulk, our results suggest a sign change while increasing the number of Co layers, in contrast to $\sigma_{\rm H}$. Meanwhile, in the S-end cases, $\alpha_{\rm N}$ vanishes for the bilayer, while it takes small negative values for the monolayer and trilayer. The layer number dependences for both cases are summarised in Fig.~\ref{fig:summary}b.

\noindent
{\bf Coulomb interactions.} Let us comment on the effect of Coulomb interactions. We tested the GGA+$U$ calculation for the S-end monolayer and bilayer systems. The results indicate that the S-end monolayer maintains the metallic state with modifications of the band structure, while the S-end bilayer shows a transition to the FM state around $U \sim 2$~eV (see Supplementary Note 4). However, in the bulk Co shandite, the results using the GGA+$U$ method are inconsistent with experimental findings~\cite{YXu}, suggesting that this method is not be suitable for this compound. Instead, a renormalisation of the band structure obtained by the GGA method explains well the ARPES measurements~\cite{DFLiu} and optical responses~\cite{YXu,Okamura}. Given the smooth connection to the bulk in the Sn-end systems, we expect that this holds also for the thin films studied here and our GGA results would be useful for comparison with experiments. Further investigation on the effect of Coulomb interactions is left to future research.

\section*{Conclusions}

Our systematic {\it ab initio} study for the thin films of Co-based shandite revealed that the number of Co kagome layers and the surface termination are both relevant to the lattice structure, the magnetic state, the electronic band structure and its topology, and transport properties. We found that the Sn-end series shows similar behaviours to the bulk, and retains the large AHE down to the monolayer limit where the quantum AHE is realised, reflecting the systematic change of the gapped Weyl nodes in the band structure. In contrast, the S-end series exhibits nonmonotonic behaviours with respect to the layer number, owing to the dominantly antiferromagnetic interlayer coupling. Our results would stimulate further experimental studies of the thin films of Co-based shandite to realise the quantum AHE in the monolayer form. The important implication of our results is that the measurements of the AHE and ANE as well as the magnetism will distinguish the film thickness and the surface termination, both of which are usually hard to directly identify in experiments. In particular, the sign of $\sigma_{\rm H}$, which is positive in the Sn-end films while negative in the S-end films, can be relevant measurement. Furthermore, the different sign of $\alpha_{\rm N}$ from the bulk will also be useful to measure the film thickness. Furthermore, our prediction of the systematic evolution of the topological band structure would stimulate optical measurements, which have been useful to detect the chirality and gap structure of the Weyl nodes away from the Fermi level~\cite{Ma,YXu,Okamura}.

\section*{Methods} 

\noindent
{\bf DFT calculation.} The OpenMX code~\cite{OpenMX,Ozaki} was used for structural optimisation and electronic structure calculations based on the density functional theory (DFT). The exchange-correlation functional was considered within the generalised gradient approximation (GGA) proposed by Perdew-Burke-Ernzerhof (PBE)~\cite{PBE}  
and the norm-conserving pseudopotentials were used. 
The Kohn-Sham wave functions were expressed by superpositions of pseudo-atomic orbitals as a basis set, which are chosen as Co6.0S-s3p2d1, Sn7.0-s3p2d2 and S6.0-s2p2d1 for bulk and monolayers, and Co6.0S-s3p3d3, Sn7.0-s4p3d1 and S6.0-s2p2d1 for bilayers and trilayers; see Ref.~\cite{OpenMX} for the details. 
For the calculations of thin films, we prepared the slabs with the bulk in-plane lattice parameter ($a=5.358~\ang$ \cite{Li}) and inserted the vacuum space along the $c$ axis which is larger than $15~\ang$. The quasi-Newton method was used for the relaxation of the atomic positions and the lattice vectors until the residual force becomes lesser than $0.01~{\rm eV} \ang^{-1}$ per atom. The structural optimisation was performed by nonrelativistic calculations for the ferromagnetic state, except for the S-end bilayer and trilayer systems, where we took interlayer antiferromagnetic and ferrimagnetic states, respectively. Then, the electronic structure calculations were performed by the fully relativistic calculations~\cite{TH}, which incorporate the effect of spin-orbit coupling, for various magnetic states (ferromagnetic, collinear antiferromagnetic and 120-degree noncollinear antiferromagnetic for the intraplane, and ferromagnetic and antiferromagnetic for the interplane), and employ the lowest-energy solution. We set the cutoff energy for the FFT grid to 1800~Ry, and sampled the Brillouin zone with $16 \times 16 \times 1$ and $16 \times16 \times 16$ $\bm k$-points for the thin films and the bulk, respectively. We confirmed that the lowest-energy magnetic states do not change in the calculations with $24 \times 24 \times 1$ and $30 \times 30 \times 1$ $\k$-points. We estimated the difference in the formation energy~\cite{Freysoldt,SZTZ} between Sn-end and S-end thin films. The formation energy for each surface termination is given as $E_{\rm f} ({\rm Co}_{3n} {\rm Sn}_{2n\pm 1} {\rm S}_{2n}) = E_{\rm DFT} ({\rm Co}_{3n} {\rm Sn}_{2n\pm 1} {\rm S}_{2n}) - 3n E_{\rm DFT} ({\rm Co}) - (2n \pm 1) E_{\rm DFT} ({\rm Sn}) - 2n E_{\rm DFT} ({\rm S}) $, where $E_{\rm DFT}$ is total DFT energy, $n$ is the layer number and $\pm$ denotes Sn-end and S-end films, respectively. The difference in the formation energy of the $n$-layer films is calculated as $\Delta E_{\rm f} = E_{\rm DFT} ({\rm Co}_{3n} {\rm Sn}_{2n + 1} {\rm S}_{2n}) - E_{\rm DFT} ({\rm Co}_{3n} {\rm Sn}_{2n - 1} {\rm S}_{2n}) - 2 E_{\rm DFT} ({\rm Sn}).$ We here set the relative chemical potential as zero for each element; comprehensive studies considering various chemical potentials should be done more systematically elsewhere~\cite{SZTZ}. The local magnetic moments are calculated from the Mulliken population analysis. 

\noindent
{\bf Band topology and transport properties.} The Bloch functions obtained by the DFT calculations were projected to the Wannier functions within the maximal localisation procedure~\cite{MV,SMV} using the OpenMX code~\cite{WOT}. We employed Co $d$, Sn $s$ and $p$, and S $p$ orbitals. The target Bloch states were chosen by setting the outer energy window properly for each system; we confirmed that the tight-binding models constructed from the Wannier functions well reproduced the original DFT band structure within the inner energy window which was also properly chosen for each case. Using the obtained tight-binding models, the anomalous Hall conductivity $\sigma_{\rm H}$ and the anomalous Nernst conductivity $\alpha_{\rm N}$ were calculated based on the Kubo formula: 

\begin{align}
\sigma_{\rm H} &= -\frac{e^2}{\hbar V_d} \sum_{{\bm k} \in {\rm BZ}} \sum_{n} f(E_{n {\bm k}}) \Omega_{n {\bm k}},
\label{eq:AHE}
\\
\alpha_{\rm N} &= \frac{e k_{\rm B}}{\hbar V_d} \sum_{{\bm k} \in {\rm BZ}} \sum_{n} s(E_{n {\bm k}}) \Omega_{n {\bm k}},
\label{eq:ANE}
\end{align}

\noindent
respectively, where $k_{\rm B}$ is the Boltzmann constant, $\hbar$ is Dirac's constant, $V_d$ is a $d$-dimensional volume of the system, $f(\ve) = (\expo^{(\ve - \mu)/k_{\rm B} T} + 1)^{-1}$ and $s(\ve) = -f(\ve) \ln f(\ve) - (1-f(\ve))\ln (1-f(\ve))$ are the Fermi distribution function and the entropy density, respectively; the summations are taken for the wave number $\k$ within the first Brillouin zone and for the band index $n$. In Eqs.~\eqref{eq:AHE} and \eqref{eq:ANE}, $\Omega_{n \k}$ is the Berry curvature calculated by

\begin{align}
\Omega_{n \k} = -\sum_{m \neq n} \frac{2{\rm Im}  \left[ J_{x,nm} J_{y,mn} \right]}{(E_{n\k}-E_{m\k})^2}, 
\end{align}

\noindent
with the current operator 

\begin{align}
J_{\xi,nm} = \left\langle n \k \left\vert \frac{\partial H_\k }{ \partial k_\xi } \right\vert m \k \right\rangle,  
\end{align}

\noindent
where $m$ and $n$ are the band indices, $\xi=x, y$, $H_\k$ is the tight-binding Hamiltonian obtained from the maximally-localised Wannier functions, and $E_{n \k}$ and $\left. \vert n \k \right\rangle$ are the eigenvalue and the eigenvector of $H_\k$, respectively. In addition, we calculate the Chern number $C_n$ for the each band $n$ by~\cite{FHS} 

\begin{align} 
C_n = \frac{1}{2\pi} \sum_{{\bm k} \in {\rm BZ}} {\rm Im} \log U_{12} U_{23} U_{34} U_{41},
\end{align}

\noindent
where $U_{ij} \equiv \langle n \k_i \vert n \k_j  \rangle$ is calculated at each plaquette on the equally-spaced $\k$-grid in the first Brillouin zone. 
The number of the $\bm k$-mesh in the first Brillouin zone was taken as $2000 \times 2000$ for monolayers and bilayers, $2800 \times 2800$ for the S-end trilayer, $6000 \times 6000$ for the Sn-end trilayer and $360 \times 360 \times 360$ for the bulk. As for $\alpha_{\rm N}$, the temperatures are set as $k_{\rm B} T =$ 0.1, 1, 5 and 10 meV, which are chosen to be lower than the Curie temperature 130~K of the thin films with the thickness 1.3 nm (roughly correspond to the trilayer)~\cite{Ikeda1}. 

\section*{Acknowledgements} 
The authors thank M. H. N. Assadi, K. Fujiwara, K. Nomura, S. Okumura, A. Ozawa and A. Tsukazaki for fruitful discussions. 
The parts of the calculation have been done using the facilities of the Supercomputer Center, the Institute for Solid State Physics, the University of Tokyo. 
This work is supported by JST CREST (Grant No. JPMJCR18T2) and JSPS KAKENHI (Grant No. JP21K13875 and JP22K03509).

\newpage



\newpage

\onecolumngrid
\clearpage
\begin{center}
\textbf{\large Supplementary Information for  \lq\lq One, two, three, $\ldots$ infinity: topological properties of thin films of Co-based shandite''}
\end{center}
\setcounter{equation}{0}
\setcounter{figure}{0}
\setcounter{table}{0}
\setcounter{page}{1}
\makeatletter
\renewcommand{\theequation}{S\arabic{equation}}
\renewcommand{\figurename}{Supplementary Figure}
\renewcommand{\tablename}{Supplementary Table}


\section*{Supplementary Note 1: Details of the optimized lattice structures} 

We present additional information of the optimised lattice structures in Supplementary Table~\ref{tab:opt_lattice}: the averaged angles for neighbouring Co-Sn-Co and Co-S-Co bonds, $\theta_{\rm Co\hy Sn\hy Co}$  and $\theta_{\rm Co\hy S\hy Co}$, respectively, and the averaged bond lengths of the nearest-neighbour Co ions in different layers, $d_{\rm inter\hy Co}$. In most cases, the bond angles and lengths do not change so much from the bulk values, except for $\theta_{\rm Co\hy S\hy Co}$ at the surfaces.

\begin{table}[h]
\centering
\setlength{\height}{3mm}
\begin{center}
\fontsize{9pt}{13pt}\selectfont
	\begin{tabular}{cclllcc} \hline 
\multicolumn{2}{c}{ System } 						&   $\theta_{\rm Co\hy Sn\hy Co}$ (degree)	& $\theta_{\rm Co\hy S\hy Co}$ (degree)	&  $d_{\rm inter\hy Co}$ $(\ang)$ 			\\ \hline \hline	
\multirow{6}{*}{Sn-end} & monolayer				&	58.8 (A-I, A-II)						&	73.6 (A-1, A-2)					&	-									\\ 
						& \multirow{2}{*}{bilayer}	&	60.4 (A-I, B-IV)						&	73.1 (A-1, B-4)					&  \multirow{2}{*}{4.69} 					\\ 
						& 						&   58.9 (A-II, B-II) 						&	77.5 (A-2, B-3)					&										\\
						& \multirow{3}{*}{trilayer}	&   60.6 (A-II, C-IV)	 					&	73.3 (A-1, C-6)					&  \multirow{3}{*}{4.67}					\\
						&						& 	59.0 (A-II, C-III)						&	77.6 (A-2, C-5)					&										\\
						& 						&	59.6 (B-II, B-III)						&	76.4 (B-3, B-4)					&										\\ \hline
\multirow{7}{*}{S-end} 	& monolayer 			&	-									&	72.0 (A-1, A-2)					&   -									\\ 	 	
						& \multirow{2}{*}{bilayer}	& \multirow{2}{*}{58.1 (A-II, B-II)}			&	71.9 (A-1, B-4)					&  \multirow{2}{*}{4.65}					\\
						& \multirow{5}{*}{trilayer}	& 						 				&	76.9 (A-2, B-3)					&  \multirow{5}{*}{4.66}					\\
						&						& \multirow{2}{*}{58.5 (A-II, C-III)}			& 	71.3 (A-1, C-6)					& 					\\[-6.5mm]			\\
						&						&										& \multirow{2}{*}{78.7 (A-2, C-5)}		&					\\[-5.5mm]			\\ 
						& 						& \multirow{2}{*}{58.5 (B-II, B-III)}			&		 							&					\\[-5.5mm]			\\	 	
						&						&										& 75.4 (B-3, B-4)					&					\\[-3.7mm]			\\ \hline
\multicolumn{2}{c}{bulk}							&	60.0		 							& 76.1								&   4.64									\\ \hline
	\end{tabular}
	\\
\vspace*{2mm}
\caption{
Additional information of the optimised lattice structures to Table~1 in the main text. $\theta_{\rm Co\hy Sn\hy Co}$ and $\theta_{\rm Co\hy S\hy Co}$ denote the averaged angles for neighbouring Co-Sn-Co and Co-S-Co bonds, respectively, and $d_{\rm inter \hy Co}$ is the averaged bond length for the nearest-neighbour Co ions in different layers, depicted in Supplementary Figure~\ref{fig:Lattice_ang}a. Alphabets, Roman numbers and Arabic numbers in the parentheses specify the Co, Sn and S ions, respectively, depicted in Supplementary Figure~\ref{fig:Lattice_ang}b. The bonds in the same parenthesis are equivalent under the $C_2$ rotation around the $\bm a$ or $\bm b$ axis. 
}
\label{tab:opt_lattice}
\end{center}
\end{table}

\begin{figure}[h]
\vspace{7mm}
\begin{center}
 \includegraphics[width=160mm]{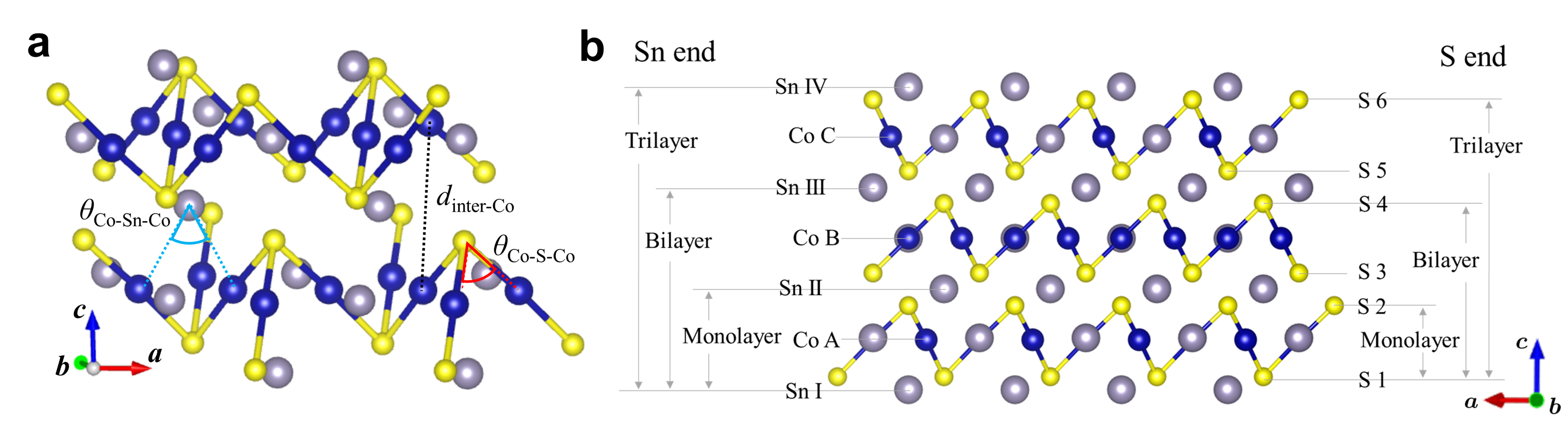}
\end{center}
\vspace{-5mm}
 \caption{
Definitions of the bond angles and the ions used in Supplementary Table~\ref{tab:opt_lattice}: 
{\bf a,} $\theta_{\rm Co\hy Sn\hy Co}$ and $\theta_{\rm Co\hy S\hy Co}$, and {\bf b,} the labels of the ions to specify the bonds for the monolayer, bilayer and trilayer cases. 
}
\label{fig:Lattice_ang}
\end{figure}

\clearpage

\section*{Supplementary Note 2: Energy comparison for various magnetic states} 

We show in Supplementary Table~\ref{tab:energy} the energy for different magnetic states obtained by the fully relativistic {\it ab initio} calculations for the optimised lattice structures. As listed in Supplementary Figure~\ref{fig:Mag}, we took the initial states as ferromagnetic (FM) states, 120 degree noncollinear antiferromagnetic (120AFM) states, and intralayer FM and interlayer collinear antiferromagnetic (IAFM) states [interlayer collinear ferrimagnetic (IFRM) state for trilayer]. For the FM states, we considered three types whose magnetic moments are along the out-of-plane ($c$) and two in-plane ($a$ and $b'$) directions; for the 120AFM states, we considered three types with different chirality and helicity; for the IAFM states, we considered three types similar to the FM cases. For the Sn-end systems, the FM states with the out-of-plane magnetic moments are the most stable, similar to the bulk, with the energy differences from the in-plane FM states being about 1$\sim$2~meV. In contrast, for the S-end systems, the stable magnetic states and the directions of the magnetic moments depend on the layer number as described in the main text. The energy differences between the most stable states and the other competing states are less than $1$~meV, which are smaller than the Sn-end cases. This implies that the magnetic anisotropy is smaller for the S-end films than the Sn-end ones. 

\begin{table}[h]
\centering
\begin{center}
\hspace*{-5mm}
\fontsize{9pt}{13pt}\selectfont
	\begin{tabular}{cccccccccc} \hline 
\multicolumn{2}{c}{\multirow{2}{*}{System}}	& \multirow{2}{*}{PM}	& \multicolumn{3}{c}{FM}	& \multirow{2}{*}{120AFM}		& \multicolumn{3}{c}{IAFM/IFRM}		\\	
					&					& 					& $a$	& $b'$	& $c$	&							& $a$	& $b'$	& $c$				\\ \hline \hline
\multirow{3}{*}{Sn-end}	& monolayer		& 122				& 0.993	& 1.01	& 0		& 85.7						& -		& -		& -					\\ 
						& bilayer		& 162				& 1.57	& 1.60	& 0		& 128						& 30.1	& 30.1	& 29.3 				\\ 
						& trilayer		& 219				& 2.28	& 2.28	& 0 		& 195						& 71.4	& 71.4	& 70.4				\\ \hline
\multirow{3}{*}{S-end} 	& monolayer 	& 393				& 0.033	& 0.034	& 0		& 162						& -		& -		& -					\\ 	 	
						& bilayer		& 79.7				& 28.6	& 28.5	& 29.5	& -							& 0.009	& 0		& 0.532				\\
						& trilayer		& 34.7				& 22.9	& 23.1	& 22.8	& -							& 0.030	& 0.046 & 0 					\\ \hline
\multicolumn{2}{c}{bulk}					& 58.1		 		& 0.619	& 0.623	& 0		& -							& -		& -		& -					\\ \hline
	\end{tabular}
	\\
\vspace*{2mm}
\caption{ 
Energy per unit cell for various magnetic states in the unit of meV. The schematic pictures of the magnetic states are shown in Supplementary Figure~\ref{fig:Mag}. The values are measured from the lowest one: FM $c$ for all the Sn-end films, and FM $c$, IAFM $b'$ and IFRM $c$ for the S-end monolayer, bilayer and trilayer, respectively. $a$, $b'$ and $c$ specify the direction of the magnetic moments. For the 120AFM states, the energy was obtained by averaging the values for three different 120 degree structures chosen such that one of the Co spin directs in the $a$, $b$ and $b'$ directions. ``-" represents the magnetic states not stable or not considered.
}
\label{tab:energy}
\end{center}
\end{table}

\begin{figure}[h]
\begin{center}
 \includegraphics[width=110mm]{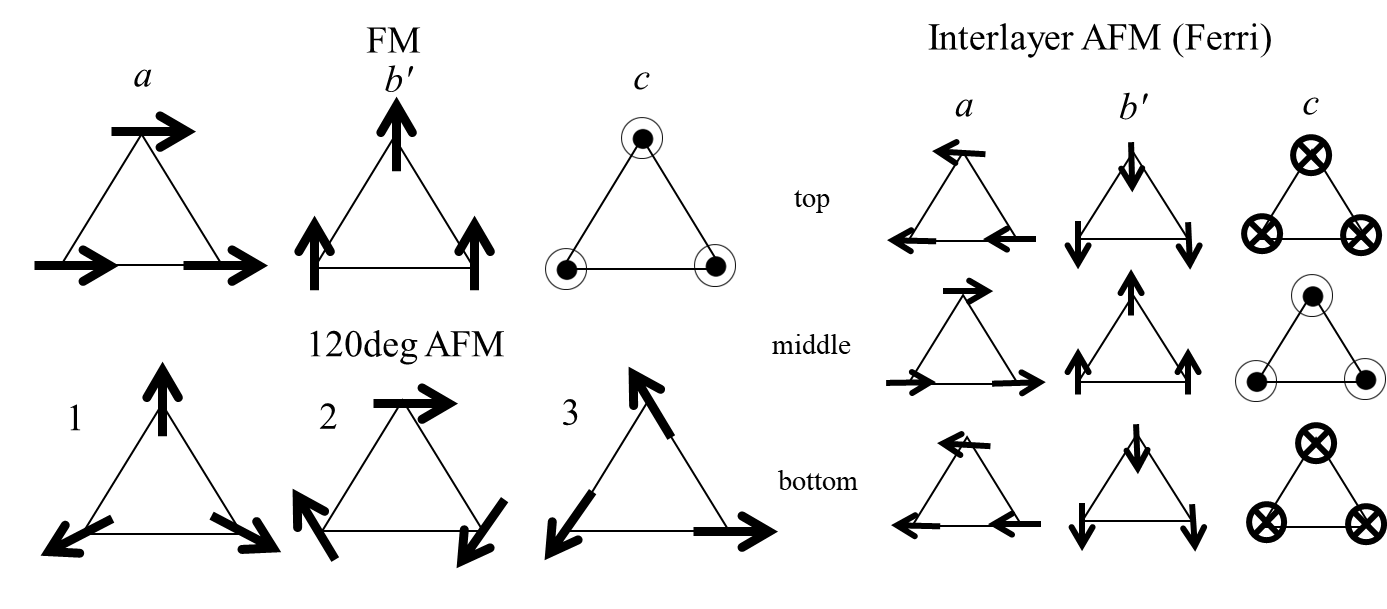}
\end{center}
 \caption{
Schematic pictures of the magnetic states considered in the {\it ab initio} calculations. The black arrows represent the directions of each Co magnetic moment on the kagome lattice. 
}
\label{fig:Mag}
\end{figure}

\clearpage

\section*{Supplementary Note 3: Maximally-localized Wannier functions for $\rm Co$ $d$ orbitals} 

We display the maximally-localised Wannier functions with Co $d_{z^2}$ and $d_{x^2 - y^2}$ orbital characters in Supplementary Figure~\ref{fig:orbitals}, which are mostly different between the Sn- and S-end films among the $d$ orbitals. 
The results are obtained for the paramagnetic state by the relativistic calculations. In all cases, the Wannier functions in the S-end systems, especially of the Co ions near the surfaces, have wider spatial extensions to the interlayer directions. This difference leads to different overlaps of the wave functions between the Sn- and S-end systems, possibly being relevant to the difference in the interlayer magnetic interactions. 


\begin{figure}[h]
\begin{center}
  \includegraphics[width=160mm]{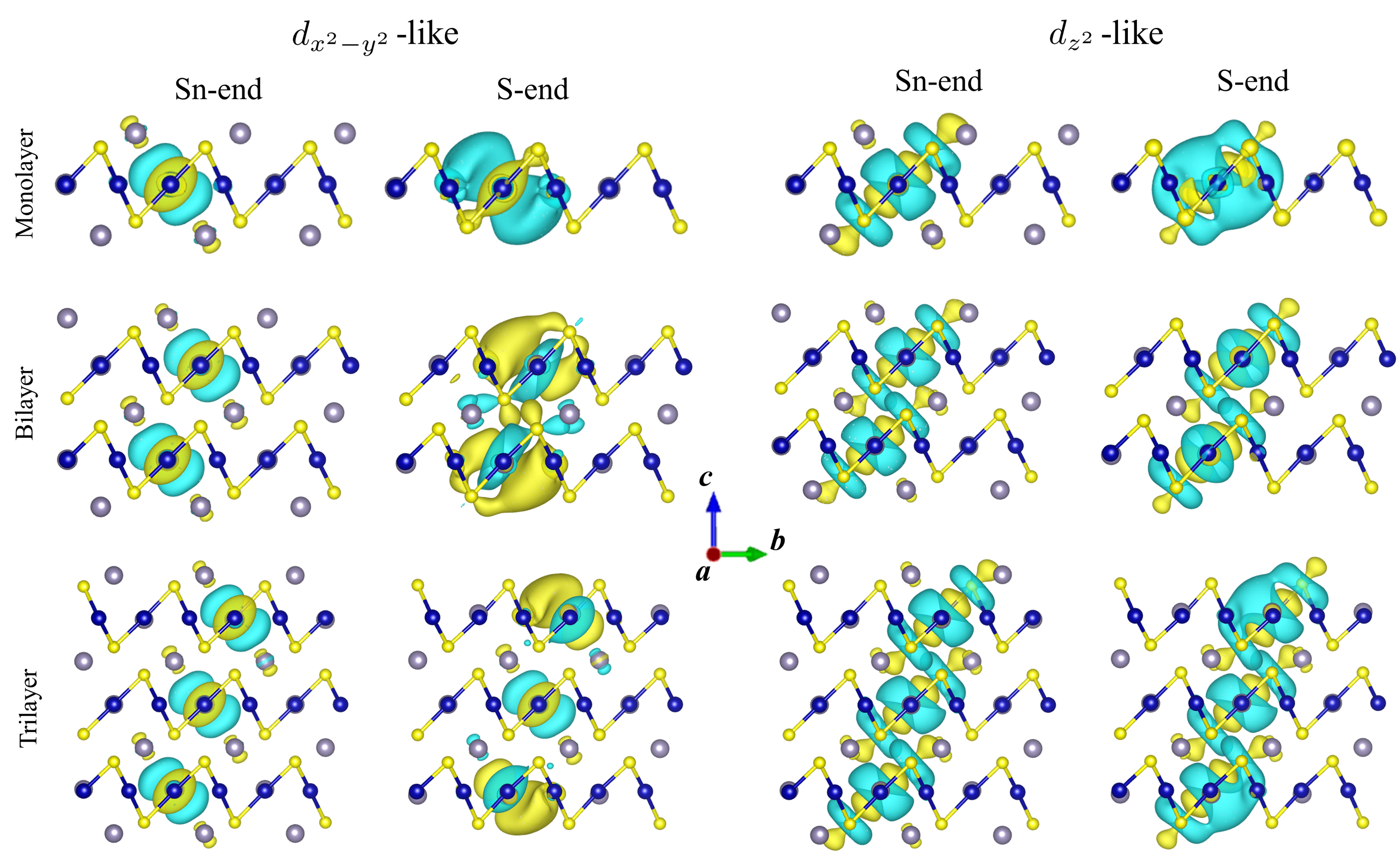}
\end{center}
 \caption{ 
 Isosurfaces of the maximally-localised Wannier functions with Co $d_{z^2}$- and $d_{x^2 - y^2}$-like orbital characters. The green and yellow surfaces represent the isosurfaces of the real part of the Wannier functions being $0.007$ and $-0.007$, respectively. 
}
\label{fig:orbitals}
\end{figure}

\clearpage

\section*{Supplementary Note 4: GGA+$U$ calculation} 

\begin{figure}[h]
\begin{center}
 \includegraphics[width=160mm]{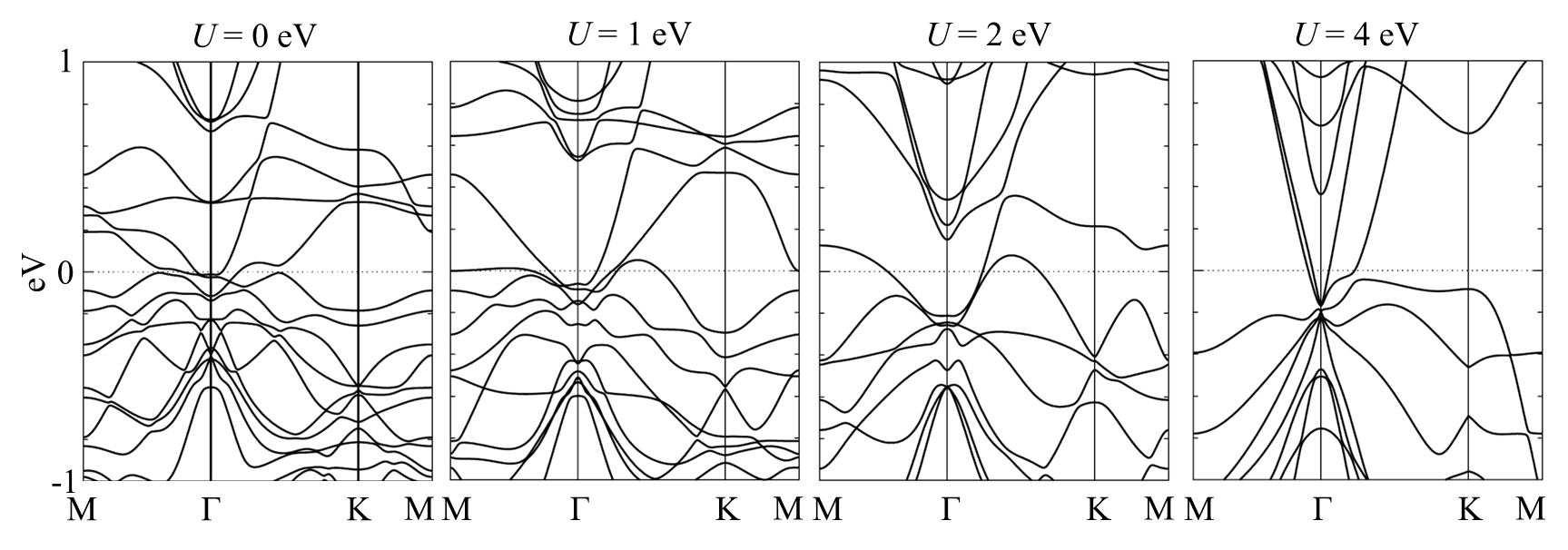}
\end{center}
 \caption{
GGA+$U$ calculations in S-end monolayer under the local Coulomb interaction on the Co $d$ electrons of $U= 0, 1, 2$ and 4 eV. 
}
\label{fig:Mono2_U}
\end{figure}

Supplementary Figure~\ref{fig:Mono2_U} shows the results of the GGA+$U$ calculations for the S-end monolayer under different $U$ on the Co $d$ electrons within the fully relativistic framework. We find that the band gap is not opened up at least to $U=4$~eV, indicating that the S-end monolayer remains metallic even when including the Coulomb interactions at the level of GGA+$U$. However, the transport properties will be altered because of the large modification of the band structure. We also performed GGA+$U$ calculations for the S-end bilayer, and find that the system turns into a FM state for $U \gtrsim 2$~eV, where the double band degeneracy is lifted and the anomalous Hall and Nernst effects are expected to occur.  

\end{document}